\documentclass[useAMS,usenatbib]{mn2e}

\title
[ Formation of  radiation pressure supported]
{Likely formation of general relativistic radiation pressure supported stars or ``eternally collapsing objects''}

\author[Abhas Mitra and Norman K. Glendenning]{Abhas Mitra$^{1}$\thanks{E-mail: amitra@barc.gov.in; NKGlendenning@lbl.gov} and Norman K. Glendenning$^{2}$\\
$^{1}$Theoretical Astrophysics Section, Bhabha Atomic Research Centre, Mumbai -400085, India\\
$^{2}$Nuclear Science Division,  Institute for Nuclear and Particle Astrophysics, Lawrence Berkeley Laboratory, Berkeley, CA. 94720}
\begin{document}
 
\date{Acccepted. Received} 
\pagerange{\pageref{}--\pageref{}} \pubyear{2010}
\maketitle
\label{firstpage}
\begin{abstract}
Hoyle \& Folwler  showed that there could be Radiation Pressure Supported Stars (RPSS) even in Newtonian gravity. 
Much later, Mitra  found  that one could also conceive  of their General Relativistic (GR) version,  ``Relativistic Radiation Pressure Supported Stars'' (RRPSSs). While RPSSs have $z\ll 1$, RRPSSs have $z \gg 1$, where $z$ is the
 surface gravitational redshift. Here we elaborate on the formation of RRPSSs during continued gravitational collapse by recalling 
  that  a contracting massive star  must   start trapping radiation as it would enter its  {\em photon sphere}.  It is found that, irrespective of the details of the  contraction process, the trapped radiation flux should attain the corresponding  Eddington value at sufficiently large $z\gg 1$.
This means that  continued GR collapse may generate an intermediate RRPSS with $z\gg 1$ before  a true BH state with $z=\infty$ is formed asymptotically. An exciting consequence of this is that the stellar mass black hole candidates, at present epoch, should be hot balls of quark gluon plasma, as has been discussed by Royzen in a recent article entitled ``{\it QCD against black holes?}''.

\end{abstract}

\begin{keywords}
gravitation -- stars: formation --stars: fundamental parameter
\end{keywords}

\section{Introduction}

For any self-gravitating object, the definition of ``compactness'' may be given in terms of the surface gravitational redshift(Weinberg 1972):
\begin{equation}
z = \left(1 - R_s/ R\right)^{-1/2} -1
\end{equation}
where $R_s = 2 G M/c^2$ is the Schwarzschild radius of the object having a gravitational mass $M$ and radius $R$. Here, $G$ is the Newtonian gravitational constant and $c$ is the speed of light. For the Sun, one has
$z \approx 2 \times 10^{-6}$ while for a typical neutron star $z \sim 0.15$. Another important parameter for a self-gravitating object is $x= p_r/p_g$,
where $p_r$ is the pure radiation pressure  and  $p_g$ is the kinetic pressure (Mitra 2009a,b). While  the central region of  Sun has $x \approx 0.006$, obviously,   a {\em cold} object
at temperature $T=0$ has $x=0$. It is known that strictly static and {\em cold} stars have an upper mass limit
both in Newtonian and Einstein gravity. In the Newtonian case, this is obtained by the marriage
of Newtonian gravity and Special Relativity, and is known as ``Chandrasekhar Mass'', $M_{ch}$. On the other hand, in Einstein gravity,
such an upper mass limit is known as ``Oppenheimer -Volkoff  Mass'' (Weinberg 1972). The precise values of such limits depend on the equation of state
(EOS) and other details. However, for a  pure {\em cold}, Helium  dwarf,   $M_{ch} \sim 1.4 M_\odot$, where
$M_\odot$ is the solar mass. On the other hand, for a pure free neutron Fermi-Dirac fluid, $M_{ov} \sim 0.8 M_\odot$. But if such upper limits were the full story, we would not have had stars of masses as large as $\sim 100 M_\odot$. Further, there are
 interstellar gas clouds of mass probably as large as $\sim 10^6 M_\odot$. The reason behind the existence of non-singular cosmic objects with such higher masses is that they are not supported by {\em cold} quantum pressure alone. On the other hand, they are supported not only by  $p_g$ but partly by  $p_r$ too (Mitra 2009a,b). 
 
  Even when one would work at a purely Newtonian level ($z \ll 1$), the probable increase in the value of $x$ would support the higher self-gravity of a star $M \gg 1 M_\odot$. And this was probably
 first realized by Hoyle \& Fowler (1963)and Fowler (1966) who conceived of the Radiation Pressure Supported Stars (RPSS) having $x\gg 1$.  It turns out that, given the self-imposed restriction, $z\ll 1$, one would require $M > 7200 M_\odot$ in order to have $x >1$ (Weinberg 1972). With the increase of $x$ and attendant self-gravity, the star tends to be more compact, i.e., $z$ would increase Mitra(2009a,b). For instance even though  ``supermassive stars'' are Newtonian (i.e., $z \ll 1$), they possess $z$ much larger than the solar value of $z_\odot \approx 2 \times 10^{-6}$. Accordingly, it is possible to conceive of  a Newtonian RPSS with $z \sim 0.1$, i.e., one which is almost as compact as a neutron star. A related important concept here is the ``Eddington Luminosity''. In order that a self-gravitating and self -luminous object
 can remain 
  quasi-static, its  luminosity 
 at a given radius $r=r$ must
 be less than a critical value (Hoyle \& Fowler 1963, Weinberg 1972):
\begin{equation}
  L_{ed}(r) = {4 \pi G M(r)
c\over \kappa}
\end{equation}
where  $M(r)$ is the gravitational mass within a given
surface, and $\kappa$ is the appropriate opacity. In case, the
star will be composed of pure ionized hydrogen, one will have the
lowest value of opacity called Thomson opacity:  $ \kappa =
\sigma_T/ m_p \approx 0.4$ cm$^2$ g$^{-1}$ where, $\sigma_T$ is the
Thomson crosssection and $m_p$ is the proton rest mass.  If one would have the intrinsic luminosity, $L >
L_{ed}$, the star would be disrupted by radiation pressure.
The Newtonian supermassive stars, conceived by Hoyle \& Fowler necessarily radiate at the above mentioned Newtonian Eddington rate. As one would
tread into Einstein gravity, however, one should be able to conceive of situations with $z \gg 1$. The definition of Eddington luminosity, in such a case, gets modified (Mitra 1998):
\begin{equation}
  L_{ed} = {4 \pi G M
c\over \kappa } (1+z)  
\end{equation}
Accordingly, in Einstein gravity, there should be  Relativistic RPSSs (RRPSS) which would {\em locally} radiate at the above mentioned enhanced Edddington rate. In the following, we discuss the basic physics behind the formation of such RRPSSs.

\section {More ~on ~Eddington ~Luminosity}
It is known that the concept of Eddington luminosity is relevant not only for the structure of stars but for the accretion process around the stars too.
 In fact,  the accretion luminosity of the star also must be
limited by $L_{ed}$ for steady spherical accretion. To appreciate
this, let us recall here the basic physics behind the concept of
``Eddington Luminosity'' by considering the fluid to be a fully
ionized hydrogen plasma: The average {\em attractive} gravitational force on one atom is $
F_g = -G M m_p/R^2$. If this would be the only force acting on the
plasma, then both the intrinsic luminosity and the accretion
luminosity  could be infinite. In reality, however,  $L$ is finite
and has a maximal value, $L_{ed}$,  because the ionized H-atom is
also subject to the {\em repulsive} force due to accretion
luminosity of the central object: \begin{equation} F_{rad} = \sigma_T ~ q/c
 \end{equation}
  where   $q = L/ 4 \pi R^2 $ is the radial energy/heat
flux. Thus the effective, \begin{equation} F_g \to F_g + F_{rad} = - {G M\over
R^2} [(1-\alpha) m_p] \end{equation}
 where $\alpha = L/L_{ed}$ and
$L_{ed} = 4 \pi G M c/ \kappa$.
Therefore, as if, the
 radially outward heat transport in a spherically symmetric
isotropic isolated body reduces the Effective  Gravity (EG)
 by a factor of $(1 -\alpha)$. This may be also seen as a reduction of the inertial mass (IM) by the same factor. 
A value of $\alpha >1$, i.e., $L > L_{ed}$ would thus mean EG to
 be negative, where either the star would be disrupted despite
 self-gravity or instead of accretion there could be radiation
 driven winds. The latter indeed happens in the atmosphere of very massive stars
 with very large $L$. In the context of collapse, if the collapse
 generated luminosity would approach its Eddington value, it becomes
 clear then that the collapse process would tend to be stalled.
Essentially, $L_{ed}$ corresponds to a critical comoving {\em outward
 heat flux} of
\begin{equation}
q_{ed} =  {L_{ed}\over 4 \pi R^2} = {G M\over \kappa R^2} (1+z)
\end{equation}
Irrespective of the specific  mode of reduction of the EG due to
heat flow, the very notion of an  ``Eddington Luminosity'', both in
Newtonian gravity and in GR, implies that {\em the attainment of}
$L=L_{ed}$ {\em would stop inflow/collapse}. 
 In fact, in a very important study on GR
collapse, Herrera \& Santos(2004) have indeed shown that outward
heat flow reduces IM by a factor $ (1 -\alpha)$.  Therefore, in principle, the GR
collapse process can certainly  {\em slow down  and get stalled}  if
the collapse generated luminosity would approach its maximal value
($\alpha \to 1$)! Further  numerical as well as analytical studies of radiative
GR collapse have
confirmed the above mentioned analytical result (Herrera \& Santos 2004;  
 Herrera, Prisco \& Barreto 2006;    
 Herrera, Prisco,  \& Ospino, 2006;  
 Herrera, Prisco, Fuenmayor, \& Troconis 2009). If such
papers (Herrera \& Santos 2004;  
 Herrera, Prisco \& Barreto 2006)     would be used for weak gravity, one would
clearly identify the $\alpha$ occurring in them as none other than
the $\alpha$ appearing in  Eq.(5).
While such studies considered enhancement of radiation flux
due to matter -radiation interaction and are somewhat non-generic, we would consider here a generic effect:  {\em the unabated enhancement of the gravitationally trapped radiation beyond the photon sphere and consequent attainment of Eddington luminosity}. 

\section {Self-Gravitational ~Trapping ~of ~Radiation}
General Relativity (GR) predicts that even the trajectories of
quanta emitted by a star itself do bend away from the direction of normal to
the direction of the tangent of  the surface of the body because of the effect of the gravitational
field of the star.    However, as long as
$z < \sqrt{3} -1$, the emitted quanta nevertheless manage to evade
entrapment and move away to infinity.
  But if the body would be so compact
as to lie within its ``photon sphere'', i.e., $R < (3/2) R_s$ or $z
> \sqrt{3} -1$, then only the radiation emitted within a cone
defined by a semi-angle $\theta_c$ (Harrison 2000):
\begin{equation}
\sin\theta_c = {\sqrt{27}\over 2} (1 - R_s/R)^{1/2} (R_s/R)
\end{equation}
will be able to escape. Radiation emitted in the rest of
the hemisphere would eventually return within the compact object.

In the presence of an external radiation, i.e., in the absence of a strict exterior vacuum, even an apparently static supermassive star
is not strictly static. This is so because, the spacetime in such a case is described by radiative Vaidya solution (Vaidya 1951) 
rather than the exact vacuum Schwarzschild solution. In this sense, while the exterior spacetime of a {\em cold} White Dwarf
or a Neutron Star having a fluid at zero tempetature is described by vacuum Schwarzschild metric, the exterior spacetime of any radiative object
including the Sun or a supermassive star, {\em in a strict sense} is described by the Vaidya metric. Accordingly, in a strict GR sense, stars are in quasistatic equilibrium and always evolving. For the Vaidya metric, the value of $z$ could be arbitrary high, and that is the reason a collapsing and radiating object is supposed to attain the black hole (BH) stage having $z=\infty$.
As the collapse/contraction proceeds, both $M$
and $R$ decrease so that $z$ increases and  eventually, the collapsing body must approach the exact BH state with $z=\infty$.
The effect of radiation trapping during these {\sl intermediate}
high-$z$ states has {\em never} been considered, though previously
 Kembhavi \& Vishveshwara(1980) observed that: 

``If neutrinos are
trapped, they will not be able to transport energy to the outside,
and this can have serious consequences on the thermal evolution of
the star. These considerations might become especially interesting
in the case of a collapsing phase which leads to the formation of a
compact, dense object.''

At high $z$, $R\approx R_s$ and from Eq.(7), one can see that, $\sin
\theta_c \to \theta_c \approx (\sqrt{27}/2) (1 +z)^{-1}$. Hence the
solid angle of escaping radiation is
\begin{equation}
\Omega_c \approx \pi \theta_c^2 \approx {27 \pi\over 4} (1+z)^{-2}
\end{equation}
The chance of escape of radiation therefore decreases as $\Omega_c/2
\pi \approx (27/8) (1+z)^{-2}$. This means that if without trapping
$10^{10}$ neutrinos/photons would escape a particular spot on the
surface, {\sl with} gravitational trapping, {\sl only} $1$ out the
$10^{10}$ quanta would escape for $z=10^5$. Consequently, as the
collapse generates internal heat/radiation, the  energy density  of
trapped radiation $\rho_r$ and associated outward heat flux {\em within} the body would increase as
\begin{equation}
q_{trap} \sim R^{-3}
(1+z)^2
\end{equation}
The   distantly observed luminosity/flux would be lesser by a factor of $(1+z)^2$  than what is indicated by the foregoing Eq. Such a reduction would take into account the fact that,
locally, trapped quanta are moving in almost closed orbits. But as far as local flux is concerned, to avoid double counting, one must not introduce any additional factor of $(1+z)$ 
in Eq.(9).
   Using Eqs.(6) and(9), we see that, in this regime of $z\gg 1$,
\begin{equation} \alpha= {q_{trap}\over q_{ed}} \sim {(1 +z)\over R M}\end{equation}
Initially, of course, $\alpha \ll 1$. But during the collapse, both $R$
and $M$ would decrease  and Eq.(10) would show that, as
$z\to \infty$, $\alpha$ would increase dramatically, $\alpha \to
1$,  at a
sufficiently high $z$. At this stage the collapse would degenerate into a secular
quasistatic contraction by the {\em very definition} of  $L_{ed}$. As if, a leaking and contracting balloon {\it stops contraction
as its self gravity fixes the leakage} by forcing  the  molecules  to move in (almost) closed circular orbits. Also, simultaneously,  the attendant heat and pressure become large enough to resist further contraction.
 In a very strict sense, however,  the body  would still be
contracting on extremely long time-scales! This is so because as
long as an horizon is not formed, i.e., $z < \infty$, the body would
radiate  and $M$ would continue to decrease. Consequently  the
metric would remain {\em non-static} and,
 in response, $R$ too, would decrease.
 It is this infinitesimal decrease in the value of $R$ and
attendant much higher secular increase in the value of $z$ and
$q_{trap}$ which would generate just  enough energy (at the expense
of $Mc^2$) to maintain the GR Eddington luminosity seen
 by a distant observer:
\begin{equation}
L^\infty_{ed} = { 4 \pi R^2 q_{ed} \over (1+z)^2} ={4 \pi G M c\over
\kappa (1+z)}
\end{equation}
Since $L^\infty = -c^2 dM/du$, the observed time scale associated with this
phase is
\begin{equation}
u = { Mc^2 \over -c^2 dM/du} = {M c^2\over L^\infty_{ed}} = { \kappa
c (1+z)\over 4 \pi G}
\end{equation}
Obviously, $u \to \infty$ {\em irrespective of the value} of $\kappa$
  as the BH stage ($z=\infty$) would be arrived.
 Thus the Eddington-limited
contracting phase actually becomes eternal. Since for  photons,  $\kappa_\gamma
\approx 0.4$ cm$^2$/g, but, for neutrinos $k_\nu$ is  smaller by an
extremely large factor of $\sim 10^{14-18}$, we will have $u_\nu \ll
u_\gamma$. Consequently, initial transition to the RRPSS phase may be
dominated by huge $\nu$-emission with a time scale $u_\nu$. But as
far as eventual secular RRPSS phase is concerned, it should be
governed by photon time scale $u_\gamma$ because it is much easier
to maintain a $L_{ed}$ caused by photons than by neutrinos:
\begin{equation}
L^\infty_{ed, \gamma} =   1.3 \left(M\over 1 M_\odot\ \right )
10^{38} (1+z)^{-1} ~{\rm erg/s}
\end{equation}

 Somewhat similar thing
happens for the formation of a hot neutron star from a proto neutron star: initial time
scale of $\sim 10$s is dictated by huge $\nu$-emission, while the
hot NS cools for thousands of years by photon emission (Glendenning 2000).
 For this era of quasi-stability by trapped photons, by using
 Eq.(13) into Eq.(12), it follows that the observed time scale of an
 RRPSS at a given $z\gg 1$ is given by
\begin{equation}
 u \approx 1.5 \times 10^{16} (1+z) ~ {\rm s} \approx 4\times 10^8
 (1+z) ~{\rm yr}
 \end{equation}
 
  For $z\gg 1$, the local energy density is almost entirely due to radiation and  pairs (Mitra 2006a) so that $\rho \approx aT^4/3$ where $a$ is the radiation constant and $T$ is the mean local temperature. Further since $M = (4\pi/3) \rho R^3$ and $R\approx 3 (M/M_\odot)$ Km for $z\gg 1$ (see Eq.[1]),  we obtain
 \begin{equation}
T = \left({3c^4\over 8\pi a G}\right)^{1/4} R_s^{-1/2} \approx 600
\left({M\over M_\odot}\right)^{-1/2}~ {\rm MeV}
\end{equation}
 Therefore, a RRPSS  is an ultrarelativistic
 fireball of radiation and  pairs interspersed
with baryons much like the plasma in the very early universe (unless $M$ is too high).  
In particular,  Eq.(15) shows that $T \sim 200$ MeV for a $10 M_\odot$ RRPSS.
 Hence, the  stellar mass RRPSSs could
be in a Quark Gluon Plasma (QGP) phase. As of now, it is believed
that a bulk QGP phase existed only in the very early universe.  But
now we arrive, through a simple and straight forward analysis, at
the exciting possibility that a {\em bulk and ever lasting QGP phase
}  may be existing within galaxies.

\section {Analytical \& Numerical Support for This Scenario}
The physical effect described here cannot be obtained by any {\em exact} analytical solution of Einstein equations simply because the {\em only exact} solution of GR gravitational collapse is the Oppenheimer - Snyder one (Oppenheimer \& Snyder 1939):
If one would consider the collapse of a homogeneous dust with $p=0$, {\em and yet  assume finite initial density}, $\rho >0$,  the fluid would appear to collapse to a singularity in a flash, $ \tau \propto \rho^{-1/2}$, with no question of slowing down or bounce or oscillation. However, physically,  a {\em strict} $p=0$ EOS should correspond to a fluid mass of $M=0$ (Ivanov 2002, Mitra 2009a) and thus, in a strict sense, a $p=0$ collapse should be eternal: $\tau =\infty$. Note, when pressure gradient forces are included, even adiabatic GR collapse {\em admits bounce and oscillatory behavior}
where the fluid need not always plunge inside its Schwarzschild radiu (Nariai 1967; Taub 1968; Bondi 1969). 
 Further Mansouri (1977) showed that a uniform
density sphere {\em cannot undergo any adiabatic collapse at all} if an equation of state would be assumed. Therefore, the effect described here here can be inferred by appropriate numerical studies of radiative physical gravitational collapse or by {\em generic} physical studies, as carried out here. Indeed realistic gravitational collapse must involve not only pressure gradient but also dissipative processes and radiation emission (Mitra 2006a,b,c; Herrera \& Santos 2004;  
 Herrera, Prisco \& Barreto 2006,    
 Herrera, Prisco,  \& Ospino, 2006;  
 Herrera, Prisco, Fuenmayor, \& Troconis 2009). 
It is clear that dissipative processes might not only slow down but even stall the collapse by generating a quasistatic state (Herrera \& Santos 2004;  
 Herrera, Prisco \& Barreto 2006,    
 Herrera, Prisco,  \& Ospino, 2006;  
 Herrera, Prisco, Fuenmayor, \& Troconis 2009). And 
 suppose, one is considering the collapse with a certain initial mass $M=M_i$ and the RRPSS state is (first) formed at $M=M_*$ and $z=z_*$. In order to find the values of $M_*$ and $z_*$, one must study the problem numerically by devising a scheme to incorporate the effect of gravitational radiation trapping and matter-radiation interaction. Obviously, the precise  value of opacity $\kappa$ would be enormously different from what has been considered here. And same would be true for all other relevant physical parameters.  But at this juncture, we are not claiming to make any such detail study. Indeed, we are interested only in a generic but physically valid picture. Implicitly, we are considering here a modest range of $M_i$  to relate it with associated numerical/analytical works (Herrera \& Santos 2004;  
 Herrera, Prisco \& Barreto 2006,    
 Herrera, Prisco,  \& Ospino, 2006;  
 Herrera, Prisco, Fuenmayor, \& Troconis 2009). On the other hand, for extremely large values of $M$, there would be no $\nu$-generation and one would be concerned with solely $\gamma$ and $e^+, e^-$ processes.

 This generic/qualitative picture becomes strengthened by the fact
Goswami \& Joshi (2005) considered the possibility that trapped surface formation may be avoided because of loss of mass by emission of radiation. Also, there is an {\em exact} solution of GR collapse which
 shows that the repulsive effects of heat flow may
prevent the formation of an event horizon (Banerjee, Chatterjee \& Dadhich 2002). Further, Fayos \& Torres (2008) have  shown that in view of the emission of radiation, GR continued collapse may turn out to be singularity free where the {\it entire mass-energy of the star may be radiated out}.  And recently, it has been found that, indeed, for radiative spherical collapse, trapped surfaces are not formed at a finite value of $M$ (Mitra 2009b).   Thus, {\em there are considerable supports from both numerical and analytical studies of Einstein equations}, about the basic feasibility of the picture presented here.

\section{Summary \& Clarifications}

  The  generic, reason why continued radiative GR collapse should (first)  result in a radiation supported quasistatic state or Eternally Collapsing Object(ECO), may be better appreciated by recalling Harrison: ``When the contracting body reaches a radius 1.5 times the Schwarzschild radius, all rays emitted tangential to the surface are curved into circular orbits. This is the radius of the {\em photon sphere}. On further contraction, the emitted rays become more strongly deflected and many now fall back to the surface. Only the rays emitted within an exit cone can escape and this escape cone narrows as contraction continues. When the body reaches the Schwarzschild radius, the exit cone closes completely and no light rays escape. Redshift and deflection conspire to ensure that no radiation escapes from a black hole.''

Just like the case of an event horizon formation is independent of the density of the fluid, {\em radiation trapping, the basic mechanism} for the scenario presented here,   too is independent of density. It simply depends on the fact for continued collapse, the fluid necessarily plunges within the {\em photon sphere} and the effect of radiation trapping must increase dramatically with increasing, $z$. Essentially we have pursued this  {\em generic} picture by recalling that there is another associated {\em generic} concept - which is
``Eddington Luminosity''. We showed that the rays falling back inside the contracting object should form an outward flux which would locally grow as $\sim (1+z)^2$ whereas the  definition of Eddington luminosity, in the increased gravity, would increase as $\sim (1+z)$.  And since  the former grows more rapidly by a factor of $(1+z)/RM$, it must {\em catch up} with the latter at some high $z=z_*\gg 1$ as the journey to the BH stage involves march towards $z\to \infty$ and $M=0$ (Mitra 2009a,b).  True, there would be many non-generic effects like matter-radiation interaction, asphericity, rotation, magnetic field etc which would affect this collapse process. But all such effects would resist the free collapse scenario and hence they would reduce the value of $z_*$. Thus this generic picture cannot be negated on the plea of complexities of the GR collapse problem.   In any case, publication of this paper may prompt researchers to
incorporate this effect of gravitational trapping in numerical studies of radiative collapse.

While the OS paper is the only {\em truly exact} study of continued gravitational collapse, it is also the most unrealistic one because it assumes $p=p_r=q=0$ even when the fluid would attain infinite density! And indeed Taub (1968) and Mansouri (1977) showed that a uniform density sphere cannot collapse at all if pressure would be incorporated! Consequently, in the past few years, several alternatives to BH Candidates have been proposed (Mazur \& Mottola 2004; Chapline 2005). Unlike the OS case, such treatments  are inexact, and further they assume that the collapsing matter with positive pressure suddendly udergoes some unspecified phase transition to acquire {\em negative} pressure. Alternately, they assume that ``Dark Energy'' starts to play dominant role at local level by virtue of mysterios quantum processes. In contrast, though, our teatment too is inexact, we simply considered  the natural GR effect that a contracting object must start trapping its own radiation once it is within its {\em photon sphere}. Note, one defines Event Horizon (EH) as a surface from which ``nothing not even light can escape''.
But the surface with $z_c = \sqrt{3} -1$ is the {\em precursor} of an EH, because  outward movement of everything gets strongly inhibited for $z > z_c$. Further, unless angular momentum is lost, everything may tend to move in closed circular  orbits for $z >z_c$. However, there must be equal number of counter rotating orbits in order to conserve angular momentum. 

In such a case, all stresses must be tangential as radial stresses should almost vanish. In other words, the situation here may approach the idealized case of an ``Einstein Cluster'' (Einstein 1939). And it follows that, when stresses are completely tangential, {\em there could indeed be static configurations with} $2M/R \to 1$ or $z \to \infty$ Florides (1974). Thus, if one would wish, one might view a RRPSS/ECO as a (quasi) static configuration with $z \gg 1$. In such a case too, there is no upper limit on the value of $M$!
There is no denying  the fact that we indeed observe compact objects, as massive as $10^{10}$ $M_\odot$. It is also certain that compact objects of even $3-4 M_\odot$s cannot be {\em cold} neutron stars. But it is fundamentally impossible to prove that these objects are true BHs simply because, by definition, {\em BH event horizons having $z=\infty$ cannot be directly detected} (Abramowicz, Kluzniak \& Lasota 2002).
And as mentioned in the introduction, for sufficiently  {\em hot} compact objects there is no upper mass limit even when stresses would be considered to be isotropic. As  we found, for stellar mass cases, the quasistatic RRPSS fluid could be in a QGP state with $T\sim 100$ MeV. Interestingly, in the context of stellar mass BH formation, a recent study entitled {\it QCD Against Black Holes} concluded that QCD phase transition would ensure that the collapsing object {\em becomes a QGP fluid rather than a true BH} (Royzen 2009).

 Note, there are many {\em observational evidences}  (Robertson \& Leiter  2002; Robertson \& Leiter, 2003; 
 Robertson \& Leiter 2004; Schild, Leiter, \& Robertson 2006; Schild, Leiter \& Robertson  2008)  for the scenario described in this paper  which show that the  compact objects in some X-ray binaries or  quasars could be  RRPSSs with $z\gg 1$ and strong intrinsic magnetic moments rather than exact BHs with $z=\infty$ and no {\em intrinsic} magnetic moment. Also, by definition, it is impossible to claim that such compact objects are true BHs with exact EHs (Abramowicz, Kluzniak \& Lasota 2002).

Unlike Mazur \& Mottola (2004) and Chapline (2005),  the scenario considered here   {\em in no way denies that,   mathematically},
the final state of continued gravitational collapse is a BH.  But, if one
would ignore the phenomenon of trapping of own radiation due
self-gravity as the body plunges into its photon sphere and also ignore the effect of pressure gradient or radiative transport, surely, one would  find  text book type prompt formation of BHs. However, the very fact that photon sphere is the precursor of an eventual EH, our scenario actually {\em fills the missing gap between a photon sphere and a true EH} during continued collapse. Accordingly, this paper may not be rejected on the assumption that it is in conflict with the basic mathematical notion of a BH.
As the {\em quasistatic} hot RRPSSs would  become more and more compact, they would asymptotically approach the ultimate state of spherical gravitational collapse; i.e., a BH with $z=\infty$ \& $M=0$ (Mitra 2009a,b). 

Newtonian and Post Newtonian radiation pressure supported stars, during their evolution, may be unstable to radial oscillation (Chandrasekhar 1965). And there is a preliminary numerical computation which suggests that  Newtonian supermassive stars may indeed collapse to form RRPSSs rather than  true BHs (Cuesta, Salim \& Santos 2005). For Newtonian supermassive stars, even though, $p_r\gg p_g$, $p \ll \rho c^2$; in contrast the RRPSSs have $p \approx (1/3) \rho c^2$. Thus  any study of Newtonian or Post Newtonian systems is not relevant for RRPSSs. Further, recently, it has been shown that, the Active Gravitational Mass Density of a quasistatic system is $\rho_g = \rho  - 3 p/c^2$ (Mitra 2010). Consequently, $\rho_g \ll \rho$ for RRPSSs and this prevents sudden rapid gravitational contraction. However, all RRPSSSs are assymptotically contracting towards the true BH state.

 To conclude, for continued collapse, we elaborated here on a most natural physical mechanism by which one may have GR version of Radiation Pressure Supported stars first conceived by Hoyle \& Fowler way back in 1963.

\section{Acknowledgements}
AM thanks Felix Afaronian for his hospitality and encouragement during a visit to the Max Planck Institute for Nuclear Physics in Heidelberg, where this work was originally conceived.

  \label{lastpage}

\begin{thebibliography}{99}
\bibitem[\protect\citeauthoryear{Abram}{2002}]{b33} Abramowicz M.A., Kluzniak W., Lasota J.P., 2002, Astron \& Astrophy., 396, L31 
\bibitem[\protect\citeauthoryear{BCD}{2002}]{b26} Banerjee A., Chatterjee S., Dadhich N., 2002, Mod. Phys. Lett.A, 17, 2335 (gr-qc/0209035)
\bibitem[\protect\citeauthoryear{Bondi}{1969}]{b22} Bondi H., 1969, MNRAS, 142, 333 
\bibitem[\protect\citeauthoryear{Chapline}{2005}]{b30} Chapline G., 2005, astro-ph/0503200
\bibitem[\protect\citeauthoryear{Chandrasekhar}{1965}]{b39} Chandrasekhar S., 1965, Astrophys. J., 142, 1519 
\bibitem[\protect\citeauthoryear{Cuesta, Salim \& Santos}{2005}]{b40} Cuesta H.J.M., Salim J.M., Santos N.O., 2005, in 100 Years of Relativity, Sao Paulo, Brazil
(see http://www.biblioteca.cbpf.br/apub/nf/NF-2005.html)
\bibitem[\protect\citeauthoryear{Einstein}{1939}]{b31} Einstein A., 1939 Ann. Math., 40, 922 
\bibitem[\protect\citeauthoryear{FayTor}{2008}]{b27} Fayos F., Torres R., 2008, Class. Q. Grav., 25, 175009 
\bibitem[\protect\citeauthoryear{Florides}{1974}]{b32} Florides P.S., 1974 Proc. Roy. Soc., London, A337, 529 
\bibitem[\protect\citeauthoryear{Fowler 1966}{1966}]{b6} Fowler W.A., 1966, Astrophys. J., 144, 180 
\bibitem[\protect\citeauthoryear{Glendenning}{2000}]{b15} Glendenning N.K., 2000 Compact Stars, Springer, NY
\bibitem[\protect\citeauthoryear{Goswami \& Joshi}{2005}]{b25} Goswami R., Joshi P.S., 2005, gr-qc/0504019 
\bibitem[\protect\citeauthoryear{Harrison}{2000}]{b12} Harrison, E., 2000, Cosmology,  The Science of the Universe, Cambridge Univ. Press, Cambridge (see pp.250)
\bibitem[\protect\citeauthoryear{Herrera \& Santos}{2004}]{b8} Herrera L., Santos N.O., 2004, Phys. Rev. D70, 084004 
\bibitem[\protect\citeauthoryear{HerPrisBar}{2006}]{b9} Herrera L., Prisco A. Di, Barreto W., 2006,  Phys. Rev. D73, 024008  
\bibitem[\protect\citeauthoryear{HerPrisOs}{2006}]{b10} Herrera L., Prisco A. Di, Ospino J., 2006, Phys. Rev. D74, 044001 
\bibitem[\protect\citeauthoryear{HerPrisFeu}{2009}]{b11} Herrera L., Prisco A. Di, Fuenmayor E., Troconis O., 2009, Int. J. Mod. Phys. D, 18(01), 129
\bibitem[\protect\citeauthoryear{Hoyle \& Folwler}{1963}]{b1} Hoyle F.,  Fowler W.A., 1963, MNRAS, 125, 169 
\bibitem[\protect\citeauthoryear{Ivanov}{2002}]{b17} Ivanov B.V., 2002, J. Math. Phys. 43(2), 1029
\bibitem[\protect\citeauthoryear{Kembhavi \& Vishveswara}{1980}]{b14} Kembhavi A.K., Vishveshwara C.V., 1980, Phys. Rev., D22, 2349
\bibitem[\protect\citeauthoryear{Mansouri}{1977}]{b23} Mansouri R., 1977,  Ann. Inst. Henri Poincare, 27(2), 173 
\bibitem[\protect\citeauthoryear{MazMot}{2004}]{b29} Mazur P.O., Mottola E., 2004, PNAS, 101, 9545 
\bibitem[\protect\citeauthoryear{MacSt}{1967}]{b19} McVittie G.C., Stable R.S., 1967, Ann. Ins. H. Poincare, A 7, 103
\bibitem[\protect\citeauthoryear{Mitra}{1998}]{b7} Mitra A., 1998, astro-ph/9811402 
\bibitem[\protect\citeauthoryear{Mitra}{2006a}]{b2} Mitra A., 2006a, MNRAS, 369, 492

\bibitem[\protect\citeauthoryear{Mitra}{2006b}]{b5} Mitra A., 2006b, MNRAS, Lett., 367, L66 (gr-qc/0601025)
\bibitem[\protect\citeauthoryear{Mitra}{2006c}]{b24} Mitra A., 2006c, Phys. Rev.,  D.74, 024010  (gr-qc/0605066)
\bibitem[\protect\citeauthoryear{Mitra}{2009a}]{b18} Mitra A., 2009a, J. Math. Phys., 50, 042502 (arXiv:0904.4754)
\bibitem[\protect\citeauthoryear{Mitra}{2009b}]{b28} Mitra A., 2009b, Pramana, 73(3), 615 (arXiv:0911.3518)
\bibitem[\protect\citeauthoryear{Mitra}{2010}]{b41} Mitra A., 2010, Phys. Lett. B. 685(1), 8 
\bibitem[\protect\citeauthoryear{Nariai}{1967}]{b20} Nariai H., 1967, Prog. Theor. Phys., 38, 740 
\bibitem[\protect\citeauthoryear{OS}{1939}]{b16} Oppenheimer J.R., Snyder H., 1939, Phy. Rev., 56(5), 455 
\bibitem[\protect\citeauthoryear{RL}{2002}]{b34} Robertson S.L., Leiter D., 2002, ApJ., 565, 447
\bibitem[\protect\citeauthoryear{RL}{2003}]{b35} Robertson S.L., Leiter D., 2003 ApJL, 596, L203 
\bibitem[\protect\citeauthoryear{RL}{2004}]{b36} Robertson S.L., Leiter D., 2004, MNRAS, 350, 1391
\bibitem[\protect\citeauthoryear{Royzen}{2009}]{b3} Royzen I., 2009, arXiv:0906.1929 (hep-th) 
\bibitem[\protect\citeauthoryear{SLR}{2006}]{b37} Schild R.E., Leiter, D.J., Robertson S.L., 2006, AJ. 132, 420 
\bibitem[\protect\citeauthoryear{SLR}{2008}]{b38} Schild R.E., Leiter, D.J., Robertson S.L., 2008 AJ., 135(3), 947 
\bibitem[\protect\citeauthoryear{Taub}{1968}]{b21} Taub A.H., 1968,  Ann. Inst. H. Poincare, 9(2), 153 
\bibitem[\protect\citeauthoryear{Vaidya}{1951}]{b13} Vaidya P.C., 1951, Proc. Ind. Acad. Sc., A33, 264
\bibitem[\protect\citeauthoryear{Weinberg}{1972}]{b4} Weinberg S., 1972, Gravitation and Cosmology, John Wiley, New York

\end{thebibliography}
\end{document}